# Can large scintillators be used for solar-axion searches to test the cosmological axion-photon oscillation proposal?

F.T. Avignone III [1], R, J. Creswick [1], and S. Nussinov [1,2,3]

[1] Department of Physics and Astronomy, University of South Carolina
Columbia, South Carolina 29208, USA

[2] Department of Physics, Tel Aviv University, Tel Aviv, Israel

[3], Physics Department, Schmid College of Science, Chapman University
Orange, California 92866

Solar-axion interaction rates in NaI, CsI and Xe scintillators via the axio-electric effect were calculated. A table is presented with photoelectric and axioelectric cross sections, solar-axion fluxes, and the interaction rates from 2.0 to 10.0 keV. The results imply that annual-modulation data of large NaI and CsI arrays, and large Xe scintillation detectors, might be made sensitive enough to probe coupling to photons at levels required to explain axion-photon oscillation phenomena proposed to explain the survival of high-energy photons traveling cosmological distances. The DAMA/LIBRA data are used to demonstrate the power of the model-independent annual modulation due to the seasonal variation in the earth sun distance.

## 1. Introduction

The CP-violating nature of QCD predicts a neutron electric-dipole moment many orders of magnitude larger than the experimental bounds. To address this problem, Peccei and Quinn postulated a global U(1) symmetry which is broken at a high energy scale [1]. The result of this spontaneous symmetry breaking is a Goldstone boson, the axion [2,3]. For about 30 years experimentalists have been searching for this particle to confirm this elegant solution of the strong-CP problem. See for example, G.G. Raffelt [4] and Hagman et al., [5], for many references to that long history. Experiments are motivated by the fact that theoretically axions can couple directly to hadrons, to electrons, or to photons. One of the main focuses here will be to address the question: could the interaction rate of solar axions, in perhaps improved large arrays of NaI and CsI crystals, and/or large xenon detectors, be large enough to search for axions at or below the sensitivity of present astrophysical bounds [4-6]? If so, might they be made sensitive enough to test the proposal that ultra high-energy photons travel cosmological distances and escape the opacity of the background radiation by oscillating into axions, or axion-like bosons, in the magnetic fields of the parent galaxy, and back to photons in the magnetic field of the Milky Way [7-9]? It was suggested by De Angelis, Roncadelli and Mansutti [7] that the fact that high-energy photons can arrive from distant astrophysical sources, without being exponentially attenuated, was evidence for the existence of a light spin-zero boson coupled to photons much like the axion or axion-like particle (ALP). Later, Simet, Hooper and Serpico suggested that the conversion of photons to ALPs might occur in the magnetic fields in and around the originating gamma-ray sources, and be converted back into photons in the magnetic fields of the Milky Way, which acts like a Sikivie magnetic helioscope [10]. More recently, Fairbairn, Rashda and Troitsky [9] investigated the possibility that this phenomenon might explain candidate neutral particles of energies $\geq 10^{18} eV$ from distant BL Lac type objects observed by a number of cosmic ray detectors. Reference [9] gives an extensive update of the experimental as well as theoretical issues.

Accordingly, in addition to the original motivations to search for axions, there is more recent interest associated with proposals that axion-photon oscillations might be a mechanism to explain the survival of TeV-photons, from very distant galaxies. This naturally motivates new interest in attempting to develop more sensitive experimental searches for ALPs. In these scenarios, photons convert to axions, or axion-like particles (ALPs), in the magnetic fields of the galaxy of origin, or in the fields of the gamma-ray emitters, via the Primakoff two-photon diagram. During a major part of their trip they remain as axions, converting back to photons in the magnetic field of the Milky Way. The well-known Lagrangian describing this process has the form:

$$L = \frac{1}{2}(\partial^\mu a \partial_\mu a - m_a^2 a^2) - \frac{1}{4} a g_{a\gamma\gamma} F_{\mu\nu} \tilde{F}^{\mu\nu} - \frac{1}{4} F_{\mu\nu} F^{\mu\nu} \ . \tag{1}$$

In equation (1), $F_{\mu\nu}$ is the electromagnetic tensor and $\tilde{F}_{\mu\nu}$ is its dual, $a$ is the pseudo-scalar axion field, and $g_{a\gamma\gamma}$ is the coupling strength of axions to photons. The middle term engenders axion-photon mixing in the presence of a magnetic field. This axion-photon oscillation is analogous to the so called axion wall experiments in which photons are converted to axions in a transverse magnetic field, pass through an opaque wall, and then through another magnetic field to convert them back to photons. In the cosmological case, the wall is the intergalactic space whose opacity is supplied by the background radiation.

In our discussions, we do not differentiate between Peccei-Quinn (PQ) axions whose coupling constants and axion masses are related by the relation $f_a m_a \propto f_\pi m_\pi$, which defines the conventional model spaces, and the general term axion to include ALPs whose coupling constants and masses are not similarly constrained. The best constraints on the axion parameters were derived from astrophysical data [5,6], and from the results of the CAST solar-axion experiment [11]. A thorough review of these bounds is given in reference [5]. It is very interesting that these bounds $\left(M \equiv g_{a\gamma\gamma}^{-1} \geq 1.1 \times 10^{10} GeV\right)$ are not very far from the coupling strength required to support the proposed axion-photon oscillation scenarios. The main question addressed in this paper is how can the discovery potential be increased to sensitivities better than the current bound, $g_{a\gamma\gamma} \approx 10^{-10} GeV^{-1}$? It will be demonstrated that large-mass scintillators with very low back ground levels give some hope of achieving this goal. However, experiments completed thus far are background limited.

We assume that axions are generated in the Sun via the Primakoff diagram describing the coupling to a two-photon vertex, very similar to the coupling of $\pi_0$ to photons. The dense flux of photons in the solar core, interacting with the Coulomb fields of nuclei, convert photons to axions via the Lagrangian:

$$L = \frac{g_{a\gamma\gamma}}{4} a F^{\mu\nu} F^{\alpha\beta} \varepsilon_{\mu\nu\alpha\beta} \ . \tag{2}$$

Equation (2) leads to the following differential interaction cross section [12]:

$$\frac{d\sigma}{d\Omega} = \frac{g_{a\gamma\gamma}^2}{32\pi^2} F_a^2(2\theta) \sin^2(2\theta) \tag{3}$$

In equations (1,2 and 3) $g_{a\gamma\gamma}$ is the coupling constant of axions to the two-photon vertex in $GeV^{-1}$. The energy spectrum of solar axions, generated by the Primakoff effect, was given by van Bibber et al. [13]. At this point we choose to express couplings to both photons and electrons in terms of two Peccei-Quinn scales, $f_{a\gamma}$ and $f_{ae}$ where $f_{a\gamma} = 8.379 \times 10^6 GeV$, in the convention used by Raffelt [4], corresponds to $g_{a\gamma\gamma} = 10^{-10} GeV^{-1}$, and the flux can be expressed as follows [5]:

$$d\Phi_a(E_a)/dE_a = \left\{\frac{8.379 \times 10^6 GeV}{f_{a\gamma}}\right\}^2 6 \times 10^{10} \cdot E_a^{2.481} (e^{-E_a/1.205}) cm^{-2} s^{-1} keV^{-1}, \tag{4}$$

In equation (4), $E_a$ is the energy of the axion. Let us further assume that the interaction of the axions with the detector, and their conversions to photons, occurs via the axio-electric effect [14,15]. Cross sections calculated with the Primakoff process are much smaller, and also yield the maximum predicted event rates in the energy range where WIMP interactions are predicted.

The formalism for the axio-electric effect was given by Dimopoulos et al., [14], and later applied to a pilot experiment to search for axions generated by the Primakoff process in the solar core [15]. The relevant expressions relating the axio-electric and photo-electric effect, corrected according to reference [16], are:

$$\sigma_{ae} = \frac{\alpha_{axion}}{2\alpha_{EM}} \left(\frac{\hbar\omega}{m_e c^2}\right)^2 \sigma_{photo-electric} \ \text{and} \ , \ \alpha_{axion} = \frac{1}{4\pi} \left(\frac{2x_e' m_e c^2}{f_{ae}}\right)^2 \tag{5}$$

$$\text{where} \ x_e' \approx 1, \ \text{and} \ \alpha_{axion} = \frac{8.312 \times 10^{-8} GeV^2}{f_{ae}^2} \tag{6}$$

$$\sigma_{axio-electric} = \frac{2.18 \times 10^{-11} GeV^2 \times (E_a^2)}{f_{ae}^2} \times \sigma_{photo-electric} \tag{7}$$

In equation (7), $E_a$ is the axion energy in keV. For purposes of demonstration, we will arbitrarily set both coupling constants to the same value. Applying equation (7), the axioelectric cross sections were computed using photoelectric cross sections calculated on-line with the MUCAL program [17].

## 2. Annual Modulation of Solar Axions

In this section we use the recent results from Bernabei et al., [18] to estimate the level of sensitivity one might achieve with annual modulation using the DAMA/LIBRA results as an example. We assume for convenience that the ALP-couplings to both photons and electrons are equal i.e., $f_{a\gamma} = f_{ae} = 8.379 \times 10^6 GeV$. We chose the axio-electric effect for detection because it yields a maximum in the expected counting rate at between 5 and 6 keV, above most of the expected signals from the scattering of WIMPS from the nuclei in the detector. In addition, the solar-axion flux is proportional to $\Omega = 1/4\pi d^2$ which is a factor of 1.0688 larger in January than in June, and corresponds to a sinusoidal amplitude of 0.0344. This would result in a significant model-independent annual modulation of a signal. The expected time-dependent rate can be expressed as follows:

$$R(E) = R_{BG}(E) + \Phi \sigma N + \eta \Phi \sigma N \sin(\omega t + \delta), \qquad (8)$$

where $R_{BG}(E)$ is the background rate, $\Phi \sigma N$ is the total axion-interaction rate, $N$ is the number of molecules/kg, so the rate is that per kg of detector, and $\eta \Phi \sigma N$ is the amplitude of the modulated signal. We now refer to Fig. 9 of reference [18] which gives the experimentally derived amplitude, $S_m (keV^{-1} kg^{-1} d^{-1})$, of the sinusoidal function of the annually modulated signal reported by the DAMA/LIBRA collaboration. It is clear from figure 9 that above 5-keV the modulation attributed to Cold Dark Matter is essentially gone, as expected because the predicted signal from WIMP scattering is also expected to be very small at this energy. We can estimate from the graph shown in Fig. 9 in reference [18], that for an annual modulation opposite in phase to that expected for WIMP interactions, $S_m(5-6keV) \leq 0.005 keV^{-1} kg^{-1} d^{-1}$. From equation (8) above we conclude $\eta \Phi \sigma N \leq 0.005 keV^{-1} kg^{-1} d^{-1}$. The expected fractional amplitude of the modulation of the ALP solar flux is $\eta = 0.0344$. Accordingly, we can derive an approximate bound, $\Phi \sigma N \leq 0.145 keV^{-1} kg^{-1} d^{-1}$. The rate given in Table I below is $1.79 \times 10^{-7} keV^{-1} kg^{-1} s^{-1}$ or $1.55 \times 10^{-2} keV^{-1} kg^{-1} d^{-1}$. The bound from Fig. 9 in the DAMA/LIBRA paper is 9.36 times higher that the rate calculated with $f_{ae} = f_{a\gamma} = 8.379 \times 10^6 GeV$. Since in this case, the rate is proportional to square of each coupling constant, the crudely estimated bound one gets from this analysis, corresponds to $\sqrt{f_{ae} f_{a\gamma}} \geq 8.379 \times 10^6 GeV / \sqrt[4]{9.36} = 4.79 \times 10^6 GeV$. This value of $f_{a\gamma}$ corresponds to $g_{a\gamma\gamma} = 1.75 \times 10^{-1} GeV^{-1}$ which clearly does not yet reach the level of sensitivity of astrophysical bounds, however the background in the DAMA/LIBRA data, ~1/keV/kg/d, is rather high, and will limit the effectiveness of this experiment for this search. We recall that the DAMA collaboration used the technique of the SOLAX collaboration [19], which takes advantage of the coherent Bragg conversion in the single crystals [20], and set a bound $g_{a\gamma\gamma} \leq 1.7 \times 10^{-9} GeV^{-1}$, corresponding to $f_{a\gamma} \geq 4.93 \times 10^5 GeV$, which is far less sensitive than that set using the annual modulation. While DAMA/LIBRA has much more data now, one can estimate that the present bound on $f_{a\gamma}$ now would not be better than a factor of two now because the rate depends on $f_{a\gamma}$ to the 4$^{th}$ power.

Very recently, the use of large Ge detector arrays was proposed to search for the 14.4 keV axion line from the M1 transition in $^{57}Fe$ in the sun [21]. The earlier work of Haxton and Lee [22], and of Moriyama [23], were used to compute interaction rates for various values of the Peccei-Quinn scale, and for various values of the flavor singlet axial vector matrix element "S". If we use the method of reference [21] we can obtain a rough bound on the couplings of axions to nucleons and to electrons, respectively. From Fig. 9 of reference [18] we can estimate that at $14.4-keV$, $\eta\Phi\sigma N \leq 0.00624$. Dividing by $\eta = 0.0344$ we obtain $\Phi\sigma N \leq 0.181 keV^{-1}kg^{-1}d^{-1}$. This is to be compared to two values corresponding to the minimum and maximum values of "$S$". They are: $3.75 \times 10^{-5}(S=0.35)keV^{-1}kg^{-1}d^{-1}$ and $1.27 \times 10^{-4}(S=0.55)keV^{-1}kg^{-1}d^{-1}$. These values correspond to $\sqrt{f_{ae}f_{a\gamma}} \geq (1.1-1.4) \times 10^6 GeV$.

While the only data available to demonstrate the potential power of this technique is from the DAMA/LIBRA experiment, there are very large xenon experiments coming on line soon with much lower background. The XENON-100 experiment is projected to have about 50-kg of fiducial mass with a background rate of less than $10^{-2} c/kg/keV/d$, and is beginning operation in the Gran Sasso Laboratory in Assergi, Italy [25]. The XMASS experiment will have a fiducial mass of 100-kg with a projected background of $10^{-4} c/kg/keV/d$ [26]. Over the next few years these two xenon experiments should provide far more sensitive annual modulation data than those used for the present demonstration.

We do not claim that any of these rough estimates are real bounds, but we use them to demonstrate the power of these large scintillators used in this way. These analyses strongly suggest the XENON-100 collaboration [25], and the XMASS collaboration [26], should analyze their future data, to search for a possible January to June modulation in solar-axion interactions in their respective detectors. It might be possible that with this technique, the bounds on the coupling of axions to photons, and to electrons, could be significantly improved. In fact, this technique could have significant discovery potential.

Table I. Axioelectric cross sections in $cm^2/molecule$, solar axion flux and predicted axion-interaction rates in $keV^{-1}kg^{-1}s^{-1}$ for $f_{PQ} = 8.379 \times 10^6 GeV$, for both interactions with photons in the sun and with electrons in the detector.

| $E_a$ keV | $\sigma_{axio-el}$ NaI | $\sigma_{axio-el}$ CsI | $\sigma_{axio-el}$ Xe | $\Phi(cm^{-2}s^{-1})$ solar – axions | $N\Phi\sigma$ NaI | $N\Phi\sigma$ CsI | $N\Phi\sigma$ Xe |
|---|---|---|---|---|---|---|---|
| 2.0 | 1.15(-42) | 2.56(-42) | 5.86(-43) | 3.18(+10) | 1.47(-7) | 1.84(-7) | 8.53(-8) |
| 3.0 | 9.16(-43) | 2.08(-42) | 4.88(-43) | 3.81(+10) | 1.40(-7) | 1.79(-7) | 8.52(-8) |
| 4.0 | 7.70(-43) | 1.80(-42) | 4.29(-43) | 3.39(+10) | 1.05(-7) | 1.38(-7) | 6.65(-8) |
| 5.0 | 1.95(-42) | 3.90(-42) | 1.11(-42) | 2.57(+10) | 2.01(-7) | 2.29(-7) | 1.31(-7) |
| 5.5 | 2.07(-42) | 6.47(-42) | 1.17(-42) | 2.15(+10) | 1.79(-7) | 3.15(-7) | 1.69(-7) |
| 6.0 | 1.96(-42) | 6.59(-42) | 1.62(-42) | 1.76(+10) | 1.39(-7) | 2.63(-7) | 1.31(-7) |
| 7.0 | 1.75(-42) | 5.97(-42) | 1.47(-42) | 1.13(+10) | 7.95(-8) | 1.53(-7) | 7.61(-8) |
| 8.0 | 1.59(-42) | 5.44(-42) | 1.35(-42) | 6.84(+9) | 4.37(-8) | 8.42(-8) | 4.23(-8) |
| 9.0 | 1.46(-42) | 5.01(-42) | 1.24(-42) | 3.99(+9) | 2.34(-8) | 4.52(-8) | 2.27(-8) |
| 10.0 | 1.35(-42) | 4.67(-42) | 1.16(-42) | 2.26(+9) | 1.22(-8) | 3.39(-8) | 1.20(-8) |

.